\documentclass [12pt]{article}

\begin{document}
\begin{center}
\textbf{Hawking radiation and entropy in de Sitter spacetime}
\end{center}

\renewcommand{\thefootnote}{\fnsymbol{footnote}}
\begin{center}Zhao
Ren$^{a}$ \footnote[2]{Corresponding author. E-mail:
zhaoren2969@yahoo.com.cn} Li Huai-Fan$^{a,b}$ Zhang Li-Chun$^{a}$ Wu
Yue-qin$^{a}$

$^{a}$Institute of Theoretical Physics, Department of Physics, Shanxi Datong University, Datong 037009 P.R.China\\
$^{b}$Department of Applied Physics, Xi'an Jiaotong University,Xi'an 710049 P.R.China\\

\end{center}

\begin{center}
\textbf{Abstract}
\end{center}

Using the analytic extension method, we study Hawking radiation of
an $(n + 4)$-dimensional Schwarzschild-de Sitter black hole. Under
the condition that the total energy is conserved, taking the
reaction of the radiation of particles to the spacetime into
consideration and considering the relation between the black hole
event horizon and cosmological horizon, we obtain the radiation
spectrum of de Sitter spacetime. This radiation spectrum is no
longer a strictly pure thermal spectrum. It is related to the change
of the Bekenstein-Hawking(B-H) entropy corresponding the black hole
event horizon and cosmological horizon. The result satisfies the
unitary principle. At the same time, we also testify that the
entropy of de Sitter spacetime is the sum of the entropy of black
hole event horizon and the one of cosmological horizon.



\section{Introduction}

Research on the correction value of B-H entropy of black holes is
one of the hot researches. There are many methods to discuss the
correction value of B-H entropy [1-18]. Most people believe that the
correction expression of B-H entropy of black hole is

\begin{equation}
\label{eq1}
S = \frac{A}{4} + \chi \ln \frac{A}{4}
 + O\left( {\frac{1}{A}} \right) + const,
\end{equation}

\noindent where $A$ is the area of the black hole horizon, $\chi $
is a dimensionless constant. However, at present, the exact value of
logarithmic term coefficient in the correction to black hole B-H
entropy is not clear.

Within the appropriate range of parameters, de Sitter spacetime not
only includes the balck hole horizon but also has the cosmological
horizon. Both the black hole horizon and the cosmological horizon
have thermal radiation. And they have different temperature. The
entropy of black hole horizon and cosmological horizon satisfy the
area formula [19]. For de Sitter spacetime, ones expect that the
thermodynamic entropy of this spacetime is $S = A_h / 4 + A_c /
4$[20,21], i.e. it is equal to the sum of the black hole horizon
entropy and cosmological horizon entropy, where $A_h $ and $A_c $
are the area of black hole horizon and cosmological horizon
respectively. At present, there is not good method to testify that
the entropy of de Sitter spacetime is the sum of the black hole
horizon entropy and cosmological horizon entropy [22].

In 2000, Parikh and Wilczek proposed the method to calculate Hawking
radiation [23]. A lot of research are made using this method, their
results all support the result obtained by Parikh and
Wilczek[24-41].

It is well-known that a black hole can be taken as a thermodynamic
system with temperature and entropy. For a spacetime that does not
include a cosmological term, all state parameters of this
thermodynamic system are reflected on the surface of the black hole
horizon. The window where the black hole transfers information to
the outside world is the black hole event horizon. However for the
spacetime that contains a cosmological term, the state parameters of
black hole not only embodies on the black hole horizon surface but
also on the cosmological horizon surface. So the window that de
Sitter spacetime transfers information to the outside world includes
the black hole horizon and the cosmological horizon. Because state
parameters of the black hole horizon and the cosmological horizon
are de Sitter spacetime state parameters, there must have relation
between the radiation of the black hole horizon and the one of the
cosmological horizon[42-44]. At present, there is not such a
research on the radiation spectrum of de Sitter spacetime after
considering the relation of the black hole horizon and the
cosmological horizon.

The organization of this work is as follows. Section 2 devotes to
extending the Damour-Ruffini method[45,46] and under the condition
that the total energy is conserved, taking the reaction of the
radiation of particles to the spacetime into consideration and
considering the relation between the black hole event horizon and
the cosmological horizon, we obtain the radiation spectrum of de
Sitter black hole. This radiation spectrum is no longer a strict
pure thermal spectrum. It is related to the change in the B-H
entropy corresponding the black hole event horizon and the
cosmological horizon. The result satisfies unitary principle. We
also derive the coefficient of the logarithmic correction term of
the B-H entropy in de Sitter black hole in section 3. We testify
that the entropy of de Sitter spacetime is the sum of the entropy of
black hole event horizon and one of cosmological horizon in section
4. Finally, we discuss our results in section 5. Throughout the
study, the units $G = c = \hbar = 1$.

\section{The radiation spectrum of $(n + 4)$-dimensional
Schwarzschild-de Sitter black hole}

We start with an $(n + 4)$-dimensional SdS black hole solution,
whose metric is [20,21]

\begin{equation}
\label{eq2}
ds^2 = - \left( {1 - \frac{r_0^{n + 1} M}{r^{n + 1}} - \lambda r^2}
\right)dt^2 + \left( {1 - \frac{r_0^n + 1}{r^{n + 1}} - \lambda r^2}
\right)^{ - 1}dr^2 + r^2d\Omega _{n + 2}^2 ,
\end{equation}

\noindent where for a positive cosmological constant $\Lambda $, we
have

\[
\lambda = \frac{2\Lambda }{(n + 2)(n + 3)},
\quad
r_0^{n + 1} = \frac{16\pi }{(n + 2)\Omega _{n + 2} },
\]
$M$ is an integration constant, $\Omega _{n + 2} $ denotes the
volume of a unit $(n + 2)$-sphere $d\Omega _{n + 2}^2 $[20,21]. The
cosmological horizon $r_c $ and black hole horizon $r_h $ are two
positive real roots of the equation, $1 - \frac{r_0^{n + 1} }{r^{n +
1}} - \lambda r^2 = 0$. The cosmological horizon is the larger one
and the black hole horizon is smaller one. The cosmological horizon
of the SdS solution has the Hawking temperature $T$ and the B-H
entropy $S$

\begin{equation}
\label{eq3}
T = \frac{1}{4\pi r_c }\left[ {\left( {n + 3)\lambda r_c^2 } \right) -
\left( {n + 1} \right)} \right],
\quad
S = \frac{r_c^{n + 2} \Omega _{n + 2} }{4}.
\end{equation}
The gravitational mass of the SdS black hole [47,48]

\begin{equation}
\label{eq4}
E = - M = \frac{r_c^{n + 1} }{r_0^{n + 1} }\left( {\lambda r_c^2 - 1}
\right).
\end{equation}

Thermodynamic quantities correspond the cosmological horizon satisfy the
first law of thermodynamics

\begin{equation}
\label{eq5}
dE = TdS.
\end{equation}
On the other hand, the black hole horizon $r_h $ in the SdS solution
has also associated Hawking temperature $\tilde {T}$ and B-H entropy
$\tilde {S}$

\begin{equation}
\label{eq6}
\tilde {T} = \frac{1}{4\pi r_h }\left[ {\left( {n + 1} \right) - \left( {n +
3)\lambda r_h^2 } \right)} \right],
\quad
\tilde {S} = \frac{r_h^{n + 2} \Omega _{n + 2} }{4}.
\end{equation}
The Abbott and Deser (AD) mass $\tilde {E}$ of the SdS solution is
[49]

\begin{equation}
\label{eq7}
\tilde {E} = M = \frac{r_h^{n + 1} }{r_0^{n + 1} }\left( {1 - \lambda r_h^2
} \right).
\end{equation}
Thermodynamic quantities correspond the black hole horizon satisfy
the first law of thermodynamics

\begin{equation}
\label{eq8}
d\tilde {E} = \tilde {T}d\tilde {S}.
\end{equation}

In curved spacetime, Klein-Gordon equation of particles with rest mass$\mu
_0 $is

\begin{equation}
\label{eq9}
\frac{1}{\sqrt { - g} }\partial _\mu (\sqrt { - g} g^{\mu \nu }\partial _\nu
\Phi ) - \mu _0^2 \Phi = 0,
\end{equation}

\noindent
where

\[
\sqrt { - g} = \sin \theta r^{n + 2}\cos ^n\theta \prod\limits_{i = 1}^{n -
1} {\sin ^i} \theta _i .
\]
Considering the condition that the radiation flow and outgoing flow
in any fixed closed supersurface should be equal, we separate
variables and derive[50]

\begin{equation}
\label{eq10}
\Phi = \frac{1}{r^{\textstyle{{n + 2} \over 2}}}e^{i\omega t}e^{im\varphi
}R(r)S(\theta )Y_{jn} (\theta _1 , \cdots ,\theta _{n - 1} ,\phi ),
\end{equation}

\noindent where $Y_{jn} (\theta _1 , \cdots \theta _{n - 1} ,\phi )$
are the hyperspherical harmonics on the $n$-sphere that satisfy the
equation

\[
\sum\limits_{k = 1}^{n - 1} {\frac{1}{\prod\nolimits_{i = 1}^{n - 1} {\sin
^i\theta _i } }} \partial _{\theta _k } \left[ {\left( {\prod\limits_{i =
1}^{n - 1} {\sin ^i\theta _i } } \right)\frac{\partial _{\theta _k } Y_{jn}
}{\prod\nolimits_{i > k}^{n - 1} {\sin ^2\theta _i } }} \right]
 + \frac{\partial _\phi \partial _\phi Y_{jn} }{\prod\nolimits_{i = 1}^{n -
1} {\sin ^2\theta _i } } + j(j + n - 1)Y_{jn} = 0.
\]

The function $R(r)$ and $S(\theta )$ are then found as the solutions to the
decoupled equations

\begin{equation}
\label{eq11}
\frac{1}{r^n}d_r \left( {r^n\Delta d_r \left( {\frac{R(r)}{r^{\textstyle{{n
+ 2} \over 2}}}} \right)} \right) + \left( {\frac{r^4\omega ^2}{\Delta } -
\Lambda _{jlm} - \mu _0^2 } \right)\frac{R(r)}{r^{\textstyle{{n + 2} \over
2}}} = 0,
\end{equation}

\begin{equation}
\label{eq12}
\frac{1}{\sin \theta \cos ^n\theta }\partial _\theta (\sin \theta \cos
^n\theta \partial _\theta S)
 + \left( {\Lambda _{jlm} - \frac{m^2}{\sin ^2\theta } - \frac{j(j + n -
1)}{\cos ^2\theta }} \right)S = 0,
\end{equation}
where $\omega $ is the energy of radiation particles, $m$ is the
projection of the radiation particle angular momentum on rotation
axis, $\Delta = r^2(1 - \lambda r^2) - r_0^{n + 1} M / r^{n - 1}$,
$\Lambda _{jlm} $ is the separation constant.

After the black hole radiated particles with energy $\omega $, $M$
in spacetime line element (\ref{eq2}) will be replaced with $M -
\omega $. Therefore, after considering the reaction of radiation to
spacetime, we define the tortoise coordinate transformation [18,51]

\begin{equation}
\label{eq13}
dr_\ast = \frac{r^2}{\Delta _\omega (r,M - \omega ,)}dr.
\end{equation}
Eq.(\ref{eq11}) is reduced to as

\begin{equation}
\label{eq14}
\frac{d^2R(r)}{dr_\ast ^2 } + \left( {\omega ^2 - \frac{\Delta _\omega
}{r^4}U(r)} \right)R(r) = 0,
\end{equation}

\noindent
where

\[
U(r) = \Lambda _{jlm} + \frac{(n + 2)(n + 3)}{4} - \frac{(n + 2)(n +
4)}{4}\lambda r^2 + \mu _0^2 r^2.
\]
Let $r_\omega $ satisfies $\Delta _\omega (r_\omega ) = 0$, thus
near $r = r_\omega $ Eq.(\ref{eq14}) can be reduced to

\begin{equation}
\label{eq15}
\frac{d^2R(r)}{dr_\ast ^2 } + \omega ^2R(r) = 0.
\end{equation}
The solution of Eq.(\ref{eq15}) is

\begin{equation}
\label{eq16}
R(r) = e^{\pm i\omega r_\ast },
\end{equation}
These solutions of ingoing wave and outgoing wave on surface $r =
r_\omega $ are respectively

\begin{equation}
\label{eq17}
\Psi _{in} = e^{ - i\omega (t + r_\ast )} = e^{ - i\omega v},
\end{equation}

\begin{equation}
\label{eq18}
\Psi _{out} (r > r_\omega ) = e^{ - i\omega (t - r_\ast )}
 = e^{ - i\omega v}e^{2i\omega r_\ast },
\end{equation}

\noindent
where $v = t + r_\ast $ is Eddington-Finkelstein coordinate. From
(\ref{eq13}), near $r = r_\omega $ we have

\begin{equation}
\label{eq19}
\ln (r - r_\omega ) = \frac{\Delta _\omega '(r_\omega )}{r_\omega ^2 }r_\ast
.
\end{equation}
Substituting (\ref{eq19}) into (\ref{eq18}), we obtain

\begin{equation}
\label{eq20}
\Psi _{out} (r > r_{\omega m} ) = e^{ - i\omega v}\left( {r - r_\omega }
\right)^{i2\omega r_\omega ^2 / \Delta _\omega '(r_\omega )}.
\end{equation}

It is obvious that the solution of outgoing wave is singular at
surface $r = r_\omega $. Eq.(\ref{eq20}) only describes the outgoing
particles outside surface $r_\omega $ and it can not describe the
outgoing particles in surface $r = r_\omega $.

According the analytic extension method proposed in
refs[18,42-46,52,53], we can obtain the outgoing wave of particle
with energy $\omega $, the outgoing rate at surface $r = r_\omega $
is given by

\begin{equation}
\label{eq21}
\Gamma _\omega = \left| {\frac{\Psi _{out} (r > r_\omega )}{\Psi _{out} (r <
r_\omega )}} \right|^2 = e^{ - 4\pi \omega r_\omega ^2 / \Delta _\omega
'(r_\omega )}.
\end{equation}
The process that the black hole radiates particles with energy
$\omega $ is an integration process [18,42-44,52,53], that is $\omega
= \int\limits_0^\omega {d\omega '} $. So the outgoing rate that the
black hole radiates particles with energy $\omega $ is

\begin{equation}
\label{eq22}
\Gamma _h (i \to f) = \prod\limits_i {\Gamma _{\omega _i } }
 = \exp \left[ { - 4\pi \int\limits_0^\omega {\frac{r_{\omega '}^2 }{\Delta
_{\omega '} '(r_{\omega '} )}d\omega '} } \right].
\end{equation}

Supposing B-H entropy difference before and after the black hole radiation
is

\begin{equation}
\label{eq23}
\Delta \tilde {S} = \tilde {S}(\tilde {E} - \omega ') - \tilde {S}(\tilde
{E}),
\end{equation}

\noindent
we have

\begin{equation}
\label{eq24}
\frac{\partial (\Delta \tilde {S})}{\partial \omega '} = \frac{\partial
\tilde {S}(\tilde {E} - \omega ')}{\partial \omega '},
\quad
d(\Delta \tilde {S}) = \frac{\partial (\Delta \tilde {S})}{\partial \omega
'}d\omega ' = \frac{\partial \tilde {S}}{\partial \omega '}d\omega ',
\end{equation}
According to the first law of thermodynamics $d\tilde {E} = \tilde
{T}d\tilde {S}$, we derive

\begin{equation}
\label{eq25}
d\tilde {S} = \frac{d\tilde {E}}{\tilde {T}},
\quad
1 / \tilde {T}(\tilde {E} - \omega ') = \frac{4\pi r_{\omega '}^2 }{\Delta
_{\omega '} '(r_{\omega '} )}.
\end{equation}
Thus

\begin{equation}
\label{eq26}
\Gamma _h (i \to f) = \prod\limits_i {\Gamma _{\omega _i } }
 = \exp \int {d(\Delta \tilde {S})} = e^{\Delta \tilde {S}},
\end{equation}

\noindent
where $\tilde {S}$ is the black hole B-H entropy.

For cosmological horizon $r_c $, using the similar method we derive
that the outgoing wave of particles with energy $\omega $ has the
following outgoing rate on cosmological horizon surface

\begin{equation}
\label{eq27}
\Gamma _c (i \to f) = \prod\limits_i {\Gamma _{\omega _i } }
 = \exp \int {d(\Delta S)} = e^{\Delta S}.
\end{equation}

\noindent where $S$ is B-H entropy corresponding cosmological
horizon.

\section{The correction to Bekenstein-Hawing entropy}

In the above calculations (\ref{eq26}) and (\ref{eq27}), we do not
consider the factor $1 / r^{(n + 2) / 2}$ in Eq.(\ref{eq10}). After
we consider this factor, from (\ref{eq21}), (\ref{eq26}) and
(\ref{eq27}) should be rewritten as

\begin{equation}
\label{eq28}
\Gamma _h (i \to f) = \frac{r_{ih}^{n + 2} }{r_{fh}^{n + 2} }e^{\Delta
\tilde {S}} = \exp \left[ {\left( {\frac{A_{fh} }{4} - \ln \frac{A_{fh}
}{4}} \right) - \left( {\frac{A_{ih} }{4} - \ln \frac{A_{ih} }{4}} \right)}
\right]
 = e^{\Delta \tilde {S}_B },
\end{equation}

\begin{equation}
\label{eq29}
\Gamma _c (i \to f) = \frac{r_{ic}^{n + 2} }{r_{fc}^{n + 2} }e^{\Delta S} =
\exp \left[ {\left( {\frac{A_{fc} }{4} - \ln \frac{A_{fc} }{4}} \right) -
\left( {\frac{A_{ic} }{4} - \ln \frac{A_{ic} }{4}} \right)} \right]
 = e^{\Delta S_B }.
\end{equation}

Thus we derive formulas for the first order correction to B-H entropy
corresponding black hole horizon and cosmological horizon.

\begin{equation}
\label{eq30}
\tilde {S}_B = \frac{A_h }{4} - \ln \frac{A_h }{4},
\end{equation}

\noindent
and

\begin{equation}
\label{eq31}
S_B = \frac{A_c }{4} - \ln \frac{A_c }{4},
\end{equation}
where $A_h $ and $A_c $ are the black hole horizon area and the
cosmological horizon area respectively. From (\ref{eq31}),
logarithmic correction term coefficient of B-H entropy in de Sitter
black hole is $ - 1$.

\section{The entropy of de Sitter spacetime}

Taking the black hole horizon and the cosmological horizon as a independent
thermodynamic system, the outgoing rate that this system radiates particles
with energy$\omega $ is obtained by Eqs.(\ref{eq28}) and (\ref{eq29}) respectively. So, the
outgoing rate that de Ditter spacetime radiates particles with energy
$\omega $ should be

\begin{equation}
\label{eq32}
\Gamma = \Gamma _h \Gamma _c = e^{\Delta \tilde {S}_B + \Delta S_B }.
\end{equation}

In quantum mechanics we obtain the probability that the system
transfers from initial state to final state is
\begin{equation}
\label{33} \Gamma (i \to f) = \left| {M_{fi} } \right|^2 \cdot
(phase \; space \; factor)
\end{equation}
The first term in the right side is the square of the amplitude, and
the phase space factor can be written as

\begin{equation}
\label{34} phase \; space \; factor = \frac{N_f }{N_i } =
\frac{e^{S_f }}{e^{S_i }} = e^{\Delta S},
\end{equation}
where $N_f $ and $N_i $ are the number of microstates of final state
and initial state of the system respectively. For a black hole, the
number of microstate is derived from final and initial
Bekenstein-Hawking entropies. So Eq.(34) is the probability that de
Sitter spacetime transfers from initial state to final state after
radiating particles with energy $\omega $ . Therefore we obtain that
the entropy of de Sitter spacetime is the sum of the entropy of
black hole event horizon and one of cosmological horizon.

\section{Conclusion}

In de Sitter spacetime both the black hole horizon and the
cosmological horizon have particle radiations. The researches on
quantum tunneling of these kinds spacetimes, Refs.[54-57] considered
that the black hole horizon and the cosmological horizon are
independent and discussed the radiation spectrums respectively. They
did not consider the relation of these two horizons.

Since de Sitter spacetime has the black hole horizon and the
cosmological horizon, the state parameters describing two horizons
are the same, the radiation of two horizons are related. To discuss
the radiation spectrum of de Sitter spacetime, we must consider the
relation of these two horizons. In this paper, we obtain the Hawking
radiation spectrum considering the relation of these two horizons
and obtain the result that the entropy of de Sitter spacetime is the
sum of the entropy of black hole event horizon and one of
cosmological horizon. This is a meaningful work for understanding
the thermodynamic properties of de Sitter spacetime with black hole
event horizon and cosmological horizon.\\
\textbf{ACKNOWLEDGMENT}

This work was supported by the Natural Science Foundation of Shanxi
Province, China (Grant No. 2006011012) and the Shanxi Datong
University
doctoral Sustentation Fund,China. \\

\end{document}